# Event-horizon-scale structure in the supermassive black hole candidate at the Galactic Centre

Sheperd S. Doeleman[1], Jonathan Weintroub[2], Alan E. E. Rogers[1], Richard Plambeck[3], Robert Freund[4], Remo Tilanus[5,6], Per Friberg[5], Lucy M. Ziurys[4], James Moran[2], Brian Corey[1], Ken H. Young[2], Daniel L. Smythe[1], Michael Titus[1], Daniel P. Marrone[7,8], Roger J. Cappallo[1], Douglas C.-J. Bock[9], Geoffrey C. Bower[3], Richard Chamberlin[10], Gary R. Davis[5], Thomas P. Krichbaum[11], James Lamb[12], Holly Maness[3], Arthur E. Niell[1], Alan Roy[11], Peter Strittmatter[4], Daniel Werthimer[13], Alan R. Whitney[1] & David Woody[12]

[1] Massachusetts Institute of Technology (MIT) Haystack Observatory, Off Route 40, Westford, Massachusetts 01886, USA. [2]Harvard Smithsonian Center for Astrophysics, 60 Garden Street, Cambridge, Massachusetts 02138, USA. [3]University of California Berkeley, Department of Astronomy, 601 Campbell, Berkeley, California 94720-3411 USA. [4]Arizona Radio Observatory, Steward Observatory, University of Arizona, 933 North Cherry Avenue, Tucson Arizona 85721-0065, USA. [5]James Clerk Maxwell Telescope, Joint Astronomy Centre, 660 North A'ohoku Place University Park, Hilo, Hawaii 96720, USA. [6]Netherlands Organization for Scientific Research, Laan van Nieuw Oost-Indie 300, NL2509 AC The Hague, The Netherlands. [7]National Radio Astronomy Observatory, 520 Edgemont Rd., Charlottesville, Virginia 22903-2475, USA[8] Kavli Institute for Cosmological Physics, University of Chicago, 5640 South Ellis Avenue, Chicago, Illinois 60637, USA. [9]CARMA, PO Box 968, Big Pine, California 93513-0968, USA. [10]Caltech Submillimeter Observatory, 111 Nowelo Street, Hilo, Hawai'i 96720, USA. [11]Max-Planck-Institut für Radioastronomie, Auf dem Hügel 69, 53121 Bonn, Germany. [12]OVRO, California Institute of Technology, 100 Leighton Lane, Big Pine, California 93513-0968, USA. [13]University of California Berkeley, Space Sciences Laboratory, Berkeley, California 94720-7450, USA.

**The cores of most galaxies are thought to harbour supermassive black holes, which power galactic nuclei by converting the gravitational energy of accreting matter into radiation[1]. Sagittarius A\*, the compact source of radio, infrared and X-ray emission at the centre of the Milky Way, is the closest example of this phenomenon, with an estimated black hole mass that is 4 million times that of the Sun[2,3]. A long-standing astronomical goal is to resolve structures in the innermost accretion flow surrounding Sgr A\* where strong gravitational fields will distort the appearance of radiation emitted near the black hole. Radio observations at wavelengths of 3.5 mm and 7 mm have detected intrinsic structure in Sgr A\*, but the spatial resolution of observations at these wavelengths is limited by interstellar scattering[4-7]. Here we report observations at a wavelength of 1.3 mm that set a size**

of $37^{+16}_{-10}$ microarcseconds on the intrinsic diameter of Sgr A*. This is less than the expected apparent size of the event horizon of the presumed black hole, suggesting that the bulk of SgrA* emission may not be not centred on the black hole, but arises in the surrounding accretion flow.

The proximity of Sgr A* makes the characteristic angular size scale of the Schwarzschild radius ($R_{sch} = 2GM/c^2$) larger than for any other black hole candidate. At a distance of ~8 kpc (ref. 8), the Sgr A* Schwarzschild radius is 10 μas, or 0.1 astronomical unit (AU). Multi-wavelength monitoring campaigns[9-11] indicate that activity on scales of a few $R_{sch}$ in Sgr A* is responsible for observed short-term variability and flaring from radio to X-rays, but direct observations of structure on these scales by any astronomical technique has not been possible. Very-long-baseline interferometry (VLBI) at 7 mm and 3.5 mm wavelength shows the intrinsic size of Sgr A* to have a wavelength dependence, which yields an extrapolated size at 1.3 mm of 20–40 μas (refs. 6,7). VLBI images at wavelengths longer than 1.3 mm, however, are dominated by interstellar scattering effects that broaden images of Sgr A*. Our group has been working to extend VLBI arrays to 1.3 mm wavelength, to reduce the effects of interstellar scattering, and to utilize long baselines to increase angular resolution with a goal of studying the structure of Sgr A* on scales commensurate with the putative event horizon of the black hole. Previous pioneering VLBI work at 1.4 mm wavelength detected Sgr A* on 980-km projected baselines, but calibration uncertainties resulted in a range for the derived size of 50–170 μas (ref. 12).

On 10 and 11 April 2007, we observed Sgr A* at 1.3 mm wavelength with a three-station VLBI array consisting of the Arizona Radio Observatory 10-m Submillimetre Telescope (ARO/SMT) on Mount Graham in Arizona, one 10-m element of the Combined Array for Research in Millimetre-wave Astronomy (CARMA) in Eastern California, and the 15-m James Clerk Maxwell Telescope (JCMT) near the summit of Mauna Kea in Hawaii. A hydrogen maser time standard and high-speed VLBI recording system were installed at both the ARO/SMT and CARMA sites to support the observation. The JCMT partnered with the Submillimetre Array (SMA) on Mauna Kea, which housed the maser and the VLBI recording system and provided a maser-locked receiver reference to the JCMT. Two 480-MHz passbands sampled to two-bit precision were recorded at each site, an aggregate recording rate of $3.84 \times 10^9$ bits per second (Gbit s$^{-1}$). Standard VLBI practice is to search for detections over a range of interferometer delay and delay-rate. Six bright quasars were detected with high signal to noise on all three baselines allowing array geometry, instrumental delays and frequency offsets to be accurately calibrated. This calibration greatly reduced the search space for detections of Sgr A*. All data were processed on the Mark4 correlator at the MIT Haystack Observatory in Massachusetts.

On both 10 and 11 April 2007, Sgr A* was robustly detected on the short ARO/SMT–CARMA baseline and the long ARO/SMT–JCMT baseline. On neither day was Sgr A* detected on the CARMA–JCMT baseline, which is attributable to the sensitivity of the CARMA station being about a third that of the ARO/SMT (owing to weather, receiver temperature and aperture efficiency). Table 1 lists the Sgr A* detections on the ARO/SMT–JCMT baseline. The high signal to noise ratio, coupled with the tight grouping of residual delays and delay rates, makes the detections robust and unambiguous.

There are too few visibility measurements to form an image by the usual fourier transform techniques, hence we fit models to the visibilities (shown in Fig. 1). We first modelled Sgr A* as a circular Gaussian brightness distribution, for which one expects a Gaussian relationship between correlated flux density and projected baseline length. The weighted least-squares best-fit model (Fig. 1) corresponds to a Gaussian with total flux density of $2.4 \pm 0.5$ Jy and full width at half maximum (FWHM) of $43^{+14}_{-8}$ µas where errors are $3\sigma$. On the assumption of a Gaussian profile, the intrinsic size of Sgr A* can be extracted from our measurement assuming that the scatter broadening adds in quadrature with the intrinsic size. At a wavelength of 1.3 mm the scattering size extrapolated from previous longer-wavelength VLBI[13] is 22 µas along a position angle 80° degrees east of north on the sky, closely aligned with the orientation of the ARO/SMT–JCMT baseline. Removing the scattering effects results in a $3\sigma$ range for the intrinsic size of Sgr A* equal to $37^{+16}_{-10}$ µas. The $3\sigma$ intrinsic size upper limit at 1.3 mm, combined with a lower limit to the mass of Sgr A* of $4 \times 10^5$ solar masses, $M_\odot$, from proper-motion work[14,15], yields a lower limit for the mass density of $9.3 \times 10^{22}$ $M_\odot$ pc$^{-3}$. This limit is an order of magnitude larger than previous estimates[7], and two orders of magnitude below the critical density required for a black hole of $4 \times 10^6$ solar masses. This density lower limit and central mass would rule out most alternatives to a black hole for Sgr A* because other concentrations of matter would have collapsed or evaporated on timescales that are short compared with the age of the Milky Way[16].

Figure 2 shows both observed and intrinsic sizes for Sgr A* over a wide range of wavelength along with the scattering model[13] and the weighted least-squares power law fit to the intrinsic size measurements. At 1.3 mm wavelength the interstellar scattering size is less than the intrinsic size, demonstrating that VLBI at this wavelength can directly detect structures in Sgr A* on event-horizon scales. The intrinsic size dependence on wavelength, $\lambda^\alpha$ ($\alpha = 1.44 \pm 0.07$, $1\sigma$), confirms that the Sgr A* emission region is stratified, with different wavelengths probing spatially distinct layers. The $\lambda^\alpha$ fit also provides an improved extrapolation to intrinsic sizes at submillimetre wavelengths consistent with emission models that produce X-ray emission from inverse Compton scattering of longer-wavelength photons[9–11]. The minimum intrinsic brightness temperature derived from our 1.3-mm results is $2 \times 10^{10}$ K.

The data presented here confirm structure in Sgr A* on linear scales of ~$4R_{sch}$, but the exact nature of this structure is not well determined. The assumption of a Gaussian model above is motivated by simplicity, but the increased angular resolution of VLBI at 1.3 mm will soon allow consideration and testing of more complex structures. As an example, the 1.3-mm VLBI data are also well fitted by a uniform thick ring of inner diameter 35 µas and outer diameter 80 µas that is convolved with the expected scattering in the interstellar medium (Fig. 1). Such structures are motivated by simulations of the Sgr A* accretion region that use full general relativistic ray tracing[17,18] and magneto-hydrodynamic effects[19], and which predict a 'shadow' or null in emission in front of the black hole position, especially in the case of face-on accretion disks. The upper limits on correlated flux density from the JCMT–CARMA baseline (Fig. 1) cannot currently discriminate between Gaussian and ring models, but expected and planned increases in both VLBI sensitivity and baseline coverage over the next 5 years will allow such detailed comparisons.

At present, Sgr A* has been shown to be coincident with the position of the unseen central mass only at the ~10 milliarcsecond level[3]. It is an open question whether or not the SgrA* source is centred on the black hole. Indeed, several models predict an offset between Sgr A* and the black hole position. In jet models of Sgr A*[20], for example, millimetre and submillimetre emission arises at a point in the relativistic plasma stream where the optical depth is close to unity, and the peak in emission can be spatially separated from the black hole. Simulations of accretion disks that are inclined to our line of sight show kinematic (Doppler) brightening on the approaching section of the disk, which also results in an emission peak that is off to one side of the black hole[17,18,19]. Even for modest accretion disk inclinations, this emission peak can be asymmetric and compact with a morphology dependant on a number of factors including black hole spin, underlying magnetic field structure, and inner disk radius.

The intrinsic size derived in this work by fitting the circular Gaussian model can be used to argue that SgrA* is not a spherically symmetric photosphere centred on the central dark mass. This is because radiation originating from a spherical surface at a given radius from a black hole is strongly lensed by gravity, and presents a larger apparent size to observers on the Earth. Such a surface of radius R centred on a non-rotating black hole will have an apparent radius, $R_a$, given by (refs. 21, 22)

$$R_a = \begin{cases} \dfrac{3\sqrt{3}}{2} R_{sch} & \text{if } R \leq 1.5 R_{sch} \\ R / \sqrt{1 - R_{sch}/R} & \text{if } R > 1.5 R_{sch} \end{cases}$$

This has the important consequence that distant observers will measure a minimum apparent diameter of ~5.2 $R_{sch}$ for all objects centred on the black hole that have radii less than 1.5 $R_{sch}$ (the minimum circular orbit for photons). In the case of SgrA*, this corresponds to a minimum apparent diameter of 52μas. This size is only marginally consistent with the 3σ upper limit on the intrinsic size derived from our 1.3mm VLBI observations, and suggests that SgrA* arises in a region offset from the black hole, presumably in a compact portion of an accretion disk or jet that is Doppler-enhanced by its velocity along our line of sight. This lensing argument also holds for the case of a maximally rotating black hole of the same mass, for which the minimum apparent size in the equatorial plane would be 45μas (ref 22), also larger than the intrinsic size derived here. The intrinsic sizes of SgrA* measured with VLBI at 3.5mm and 7mm exceed the minimum apparent size, and thus cannot similarly be used to constrain the location of SgrA* relative to the black hole.

Detection of the event-horizon scale structure reported here indicates that future VLBI observations at $\lambda \leq 1.3$ mm will open a new window onto fundamental black hole physics through observations of our Galactic Centre. Plans to increase the sensitivity of the VLBI array described here by factors of up to 10 are under way, and the addition of more VLBI stations will increase baseline coverage and the ability to model increasingly complex structures. At projected VLBI array sensitivities, Sgr A* will be detected on multiple baselines within 10-s timescales, allowing sensitive tests for time-variable structures such as those suggested by orbiting hotspot[18] and flaring models[9-11].

**Acknowledgements** High-frequency VLBI work at MIT Haystack Observatory is supported by grants from the National Science Foundation. The Submillimeter Array is a joint project between the Smithsonian Astrophysical Observatory and the Academia Sinica Institute of Astronomy and Astrophysics. We thank G. Weaver for the loan of a frequency reference from Johns Hopkins University Applied Physics Labs; J. Davis for use of GPS equipment; I. Diegel, R. Vessot, D. Phillips and E. Mattison for assistance with hydrogen masers; the NASA Geodesy Program for loan of the CARMA Hydrogen Maser; D. Kubo, J. Test, P. Yamaguchi, G. Reiland, J. Hoge and M. Hodges for technical assistance; M. Gurwell for SMA calibration data; A. Kerr and A. Lichtenberger for contributions at ARO/SMT; A. Broderick, V. Fish, A. Loeb, and I. Shapiro for helpful discussions; and the staff at all participating facilities.

Table 1 VLBI detections of Sgr A* on the ARO/SMT–JCMT baseline at 1.3 mm wavelength

| Date (UT) | Correlated flux density (Jy) | SNR | Residual delay (ns) | Residual delay rate (ps s$^{-1}$) | Projected baseline length ($10^6 \lambda$) |
|---|---|---|---|---|---|
| 10 April 2007 12:20 | 0.38 | 5.8 | −4.9 | −0.29 | 3,558 |
| 11 April 2007 11:00 | 0.37 | 5.0 | −7.2 | −0.25 | 3,443 |
| 11 April 2007 11:40 | 0.34 | 5.4 | −7.9 | −0.21 | 3,535 |
| 11 April 2007 12:00 | 0.31 | 5.8 | −8.0 | −0.19 | 3,556 |

Columns are observation date, correlated flux density on the ARO/SMT–JCMT baseline, signal to noise ratio of the VLBI detection, delay and delay-rate residual to the correlator model, and the baseline length projected in the direction of Sgr A*. Each detection was made by incoherently averaging[23] the VLBI signal and searching for a peak in signal to noise ratio over a range of ±18 ns in delay and ±2 ps s$^{-1}$ in delay rate (500 Nyquist sampled search points). For detections on 11 April, data were averaged over 10-min observing scans. The detection on 10 April averaged two 10-min scans together at 12:20 and 12:40 UT to increase integration time. The offset in residual delay between 10 April and 11 April is due to slowly varying instrumental effects and is seen at this same level for nearby quasar calibrators. The statistics of VLBI fringe detection with incoherent averaging are non-Gaussian, and the probability of false detection (the chance a pure noise spike could masquerade as a detection) is a very sharp function of SNR. In the fringe searches on the ARO/SMT–JCMT baseline, for example, SNR of 4.5 is required to give a robust probability of false detection of <$10^{-6}$, and for SNR of 5.8 in the incoherent fringe search, the probability of false detection is below $10^{-9}$. Out of a total of 15 separate 10-min scans, Sgr A* was detected four times on the ARO/SMT–JCMT baseline. Given the strength of these detections, one would expect a higher detection rate than the observed 25%. The low detection rate could be due to intrinsic variations in Sgr A* flux density, but it is more likely to be due to a combination of both pointing errors and variable atmospheric coherence, which would lower fringe search sensitivity, especially at the low elevations at which all sites observed Sgr A*. To convert to Jy, data were calibrated using system temperature, opacity and gain measurements made at all sites.

**Figure 1 Fitting the size of Sgr A* with 1.3 mm wavelength VLBI.** Shown are the correlated flux density data on the ARO/SMT–CARMA and ARO/SMT–JCMT baselines plotted against projected baseline length (errors are $1\sigma$). Squares show ARO/SMT–CARMA baseline data and triangles show ARO/SMT–JCMT data, with open symbols for 10 April and filled symbols for 11 April. The solid line shows the weighted least-squares best fit to a circular Gaussian brightness distribution, with FWHM size of 43.0 $\mu$as. The dotted line shows a uniform thick-ring model with an inner diameter of 35 $\mu$as and an outer diameter of 80 $\mu$as convolved with scattering effects due to the interstellar medium. The total flux density measurement made with the CARMA array over both days of observing ($2.4 \pm 0.25$ Jy : $1\sigma$) is shown as a filled circle. An upper limit for flux density of 0.6 Jy, derived from non-detection on the JCMT–CARMA baselines, is represented with an arrow near a baseline length of $3,075 \times 10^6 \lambda$.

**Figure 2 Observed and intrinsic size of Sgr A* as a function of wavelength.** Red circles show major-axis observed sizes of Sgr A* from VLBI observations (all errors $3\sigma$). Data from wavelengths of 6 cm to 7 mm are from ref. 13, data at 3.5 mm are from ref. 7, and data at 1.3 mm are from the observations reported here. The solid line is the best-fit $\lambda^2$ scattering law from ref. 13, and is derived from measurements made at $\lambda > 17$ cm. Below this line, measurements of the intrinsic size of Sgr A* are dominated by scattering effects, while measurements that fall above the line indicate intrinsic structures that are larger than the scattering size (a 'source-dominated' regime). Green points show derived major-axis intrinsic sizes from 2 cm $< \lambda <$ 1.3 mm and are fitted with a $\lambda^\alpha$ power law ($\alpha = 1.44 \pm 0.07$, $1\sigma$) shown as a dotted line. When the 1.3-mm point is removed from the fit, the power-law exponent becomes $\alpha = 1.56 \pm 0.11$ ($1\sigma$).

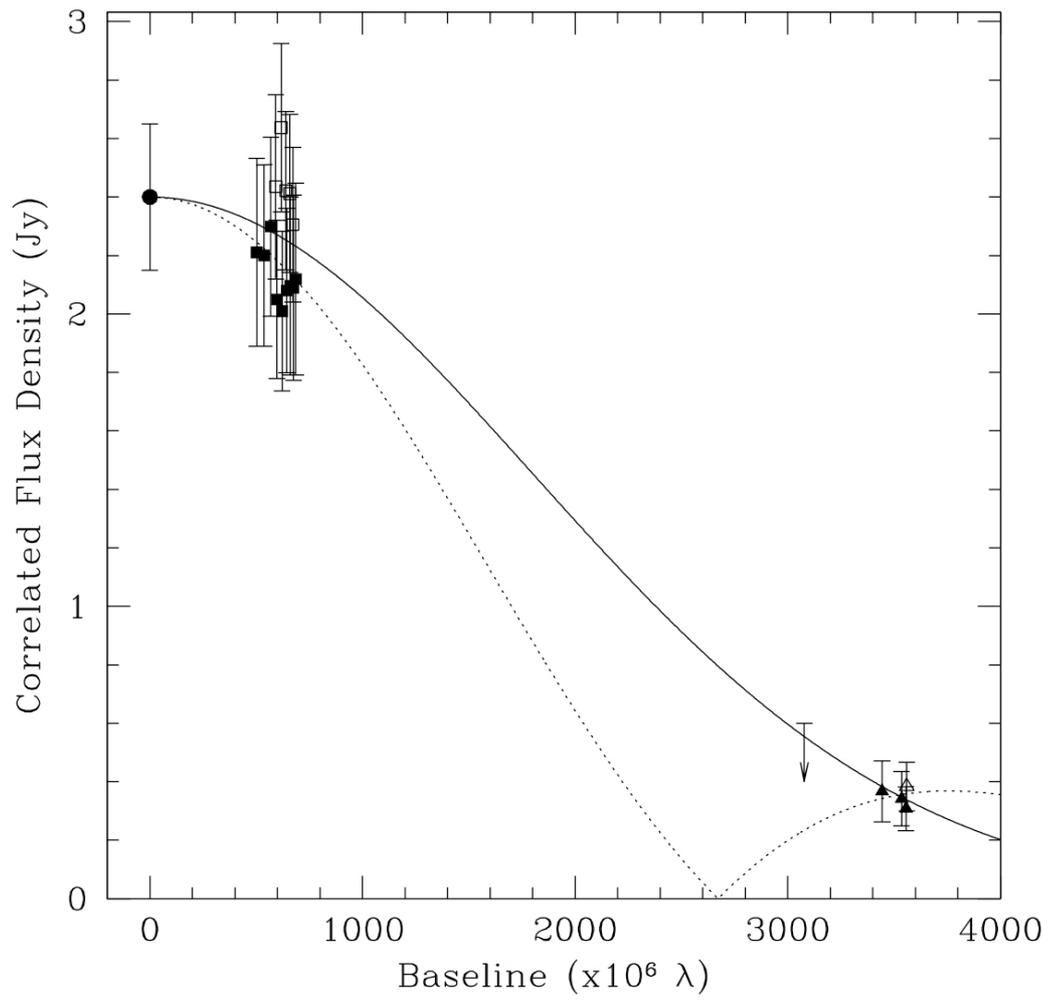

**Fig 1**

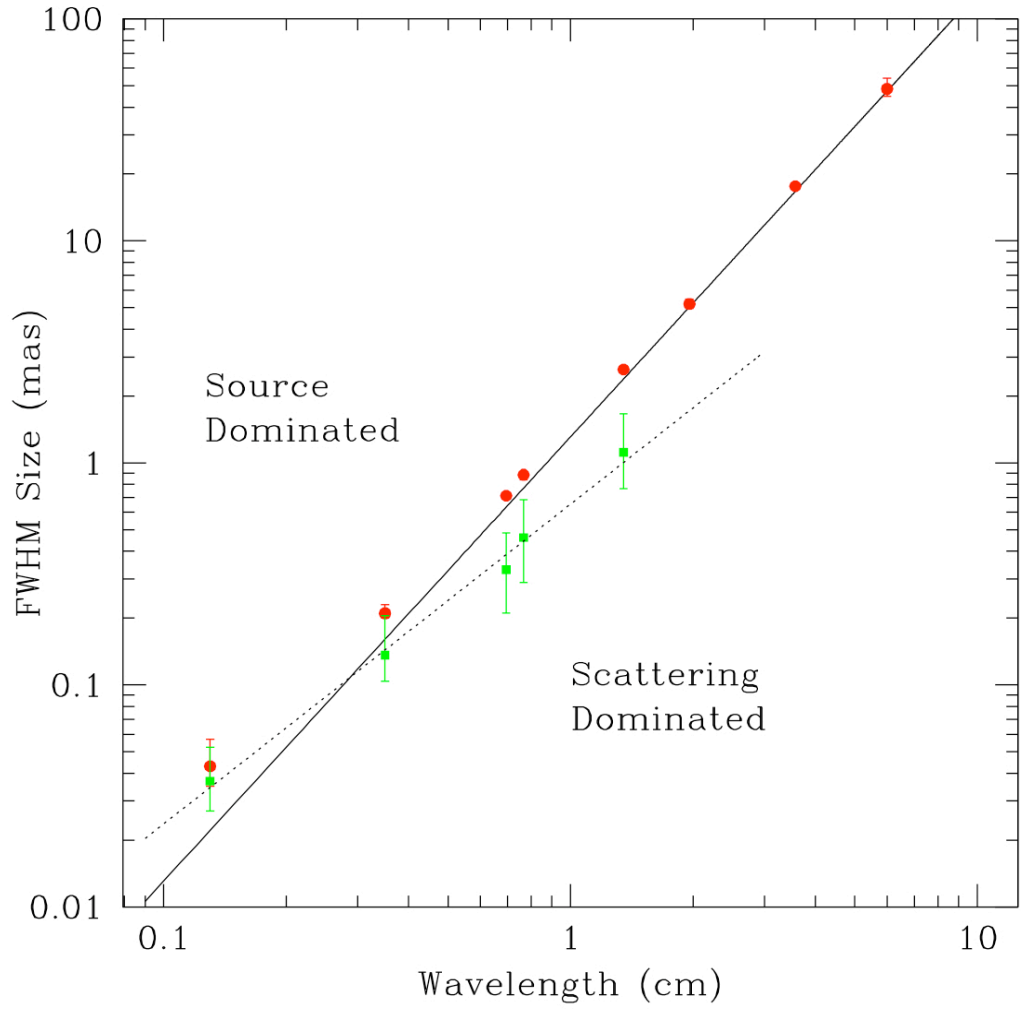

**Fig 2**